\documentclass{ws-procs9x6}

\begin{document}

\title{The quantum geometry of tensorial group field theories}

\author{D. Oriti$^*$}

\address{Albert Einstein Institute \\ Am M\"uhlenberg 1, 14476 Potsdam-Golm, Germany, EU\\
$^*$E-mail: daniele.oriti@aei.mpg.de}

\begin{abstract}
We remark the importance of adding suitable pre-geometric content to tensor models, obtaining what has recently been called tensorial group field theories, to have a formalism that could describe the structure and dynamics  of quantum spacetime. We also review briefly some recent results concerning the definition of such pre-geometric content, and of models incorporating it. 
\end{abstract}


\bodymatter

\

The last few years have witnessed a revival of tensor models \cite{tensor}, as a way to generalize to higher dimensions the successes of matrix models \cite{mm} in describing 2d quantum gravity as a theory of random surfaces. Historically, this revival started \cite{DP-F-K-R} in the area of spin foam models \cite{SF}, a covariant version of the dynamics of loop quantum gravity (LQG) \cite{LQG}. A complete definition of such dynamics was indeed proposed in the form of {\it group field theories} (GFTs) \cite{GFT}, combinatorially non-local field theories on group manifolds whose Feynman diagrams are given by d-dimensional simplicial complexes, and whose Feynman amplitudes are given by the same spin foam amplitudes encoding the quantum dynamics of spin networks states in LQG. This formulation also suggested \cite{GFTemergence} a change of perspective on the same dynamics. The spin foam approach has developed to a great extent, with the construction of new models and an increased understanding of their quantum geometric aspects \cite{SF,QG-SF-sigma}. To go beyond the truncation of degrees of freedom represented by any single simplicial complex, towards an approximately continuum physics, remains however a pressing issue. This is basically a problem in renormalization and of extracting effective dynamics from the fundamental one. One strategy is suggested by a lattice gauge theory perspective on spin foam amplitudes and involves background independent coarse graining \cite{bianca}. The other strategy uses the GFT implementation of the spin foam dynamics and standard QFT renormalization \cite{GFTrenorm} and mean field theory \cite{GFTmean} tools. This second strategy rests on the new developments of tensor models \cite{TensorReview, uncoloring}, in particular the discovery of a large-N expansion \cite{RazvanLargeN}, which have led to improved analytic control. This also opens the possibility of a better analytic control over GFTs, and thus the full dynamics of spin foam models and LQG, thanks to these tensorial tools. In order to stress this possibility we refer to them in the following as {\it Tensorial Group field Theories}. Rather than calling for a wider application of TGFTs in spin foam and LQG research, the main point we want to make in this paper is that, if the goal is to solve the problem of Quantum Gravity, we need to study interesting tensorial group field theories with their richer quantum geometric structure rather than the simpler tensor models, for which most of the analytic results have been obtained up to now.

\section{Basics of tensorial field theories}
The type of models that have been studied up to now fall into two categories. The first is {\it tensor models}, an uncolored simplicial $d=3$ example being:  
\begin{equation}
S(T) = \, \frac{1}{2} \sum_{i,j,k} T_{ijk} T_{kji} \, -\, \frac{\lambda}{4!\sqrt{N^3}}\, \sum_{ijklmn} T_{ijk} T_{klm} T_{mjn} T_{nli}
\end{equation}
where the complex tensor over $(\mathbb{Z}_N)^3$ can be graphically associated to a triangle with edges labelled by $i=1,..,N$, and the interaction has the combinatorics of the gluing of four triangles along edges to form a tetrahedron. The coloring of the same model with additional labels on the tensors could be used to define interactions with a $U(N)^d$ invariance, corresponding to more general polytopes \cite{uncoloring}.
The second is {\it Tensorial Group Field theories}, an uncolored, simplicial 3d example\footnote{Clearly, if neither coloring nor tensor invariances are used, what we have is really just an ordinary GFT, which we however treat as a special case of TGFTs.} being the Boulatov model:
\begin{eqnarray}
S_{3d}[\varphi,\bar{\varphi}]\,=\, \frac{1}{2}\int[dg]^3
\bar{\varphi}(g_1,g_2,g_3)\varphi(g_3,g_2,g_1) \;- \hspace{4cm} \nonumber \\ -\;\frac{\lambda}{4!}\int [dg]^6
\varphi(g_1,g_2,g_3)\varphi(g_3,g_4,g_5)\varphi(g_5,g_2,g_6)\varphi(g_6,g_4,g_1) \quad +\quad c.c.  \nonumber
\end{eqnarray}
where the basic variables is a (complex) field over $G^3$, with $G$ a Lie group ($SU(2)$ for the 3d quantum gravity model), assumed to possess the invariance: $\varphi(g_i) = \varphi(h g_i)$, and the same combinatorics as the simple tensor model above. The invariance property is a simple example of additional feature imposed on a tensorial field, motivated, as the choice $G=SU(2)$, by quantum geometric considerations \cite{GFT}.

All the above models are then defined, at the quantum level, by the perturbative expansion of the partition function in Feynman diagrams $\Gamma$, which correspond to arbitrary gluings of d-simplices (or other polytopes) along their (d-1)-faces. What physics one seeks to encode in the formalism depends on the exact choices of action (interaction kernel and propagator) and data associated to the basic tensorial field (domain space). This area of research has witnessed an impressive growth in recent years, with results on many aspects of the formalism: the construction of models for quantum gravity as well as statistical systems, studies on classical and quantum symmetries of the same models, analyses on the topology of the cellular complexes generated in perturbative expansion, the large-N limit, studies on perturbative TGFT renormalization, works on summability and critical behaviour, and the extraction of effective physics. For all this, we refer to the literature. Now, we focus on the definition of quantum gravity models.

\section{Pre-geometric data: phase space, quantization maps, flux representation}
The phase space underlying quantum gravity TGFT models is the cotangent bundle over a Lie group: $\mathcal{T}^*G\simeq G \times \mathfrak{g}$, with, in particular, $G=SU(2)$. This is the building block of the phase space of both simplicial gravity and LQG \cite{SF,GFT,LQG}. The group elements are interpreted as parallel transports of an $SU(2)$ connection along elementary links of a graph or of a (dual of a) simplicial complex, and the conjugate Lie algebra elements as fluxes of a dual (densitized) co-triad across (d-2)-faces dual to the same links. We refer to the literature for more details on the quantization. Here we report on some recent work \cite{CarlosDanieleMatti} on a new representation for the quantum theory \cite{GFT-NC-review}.

The fundamental poisson brackets are  
\begin{equation}
\{\zeta^i(g),\zeta^j(g)\} = 0 \ ,\ \{X_i,\zeta^j(g)\} = \tilde{\mathcal{L}}_i\zeta^j(g) \ ,\ \{X_i,X_j\} = \kappa\epsilon_{ij}^{\phantom{ij}k}X_k
\end{equation}
where $\zeta$ are coordinates on the group manifold and $\mathcal{L}$ is a Lie derivative. Any choice of {\it quantization map} $\mathcal{Q}\,: \,C^\infty(\mathcal{T}^*SU(2)) \rightarrow \mathcal{A}$ will give the corresponding algebra of operators acting on some Hilbert space $\mathcal{H}$. Given the commutativity of functions on the group, a standard basis is given by group-labelled states, which provide a realization of the Hilbert space as $L^2(G)$. One can also look for a dual realization in terms of non-commutative functions of Lie algebra elements $X\in\mathfrak{su}(2)\simeq \mathbb{R}^3$, endowed with a $\star$-product following uniquely from the quantization map:
\begin{equation}
f \star h \equiv \mathcal{Q}^{-1}(\mathcal{Q}( f ) \mathcal{Q}( h )) \qquad .
\end{equation}
This allows to define the (flux) representation
\begin{equation}
\hat{\zeta}^i \tilde{\psi}(X) = -i\hbar\frac{\partial}{\partial X_i} \tilde{\psi}(X) \quad , \quad \hat{X}_i \tilde{\psi}(X) = \tilde{\psi}(X) \star X_i \qquad .
\end{equation} 
The coefficients of the change of basis from group-labelled states to Lie algebra-labelled states are non-commutative plane waves:
\begin{equation}
E_g(X) := e_\star^{\frac{i}{\hbar\kappa}k_g\cdot X} = \mu(g) e^{\frac{i}{\hbar}\zeta(g)\cdot X} \label{planewave}
\end{equation}
$k_g$ are the coordinates obtained by inverse exponential map, and the $\star$-exponential is defined by the series expansion in $\star$-monomials of Lie algebra elements.
The plane waves are thus generically $\star$-exponentials for the $\star$-product defined from the quantization map, and can be written as standard exponentials for some choice of coordinates on the group which also follows uniquely from the quantization map. They satisfy: $\left(E_{g_1}\star E_{g_2}\right)(X) = E_{g_1 g_2}(X)$. Using them, one can then define a unitary intertwining map between group and Lie algebra representations: a non-commutative Fourier transform. The flux representations and non-commutative Fourier transforms that follow from various quantization maps have been studied in ~\cite{CarlosDanieleMatti}. The one based on the plane waves $E_g(x) = e^{i \zeta_g \cdot x}$, with $\zeta_g^i = \sin\theta n^i$, with $g= \cos\theta I + i \sin\theta n\cdot \sigma$ in the fundamental representation ($\sigma^i$ are the Pauli matrices), has already found several applications in quantum gravity \cite{GFTmetric, GFT-NC-review}.

\section{A TGFT model for 4d quantum gravity}
We now give an example of a TGFT model for 4d gravity \cite{GFT-Holst}. We aim at a description of quantum spacetime as the result of the interaction of fundamental building blocks represented by quantum tetrahedra, the quanta of our TGFT field, and at encoding appropriately their quantum geometry. A geometric tetrahedron in $\mathbb{R}^4$ can be described by four bivectors $B_i \in \wedge^2\mathbb{R}^4\simeq \mathfrak{so}(4)\simeq \mathfrak{su}(2)\oplus\mathfrak{su}(2)$, associated to its four triangles, and a vector $k\in S^3$, interpreted as its (unit) normal, satisfying:
\begin{equation}
N_I\,(*B_i^{IJ})\,=\,0  \qquad (simplicity) \qquad \sum_iB_i^{IJ}\,=\,0 \qquad (closure)
\end{equation} 
So the classical phase space for a tetrahedron, before the imposition of constraints, is $\left[\mathcal{T}^*SO(4)\right]^4\simeq \left[\mathcal{T}^*SU(2)\times\mathcal{T}^*SU(2)\right]^4$.
In selfdual and anti-selfdual components, the first condition becomes $B_+^i + k B_-^i k^{-1} = 0$, where $\bar{k}\!:=\!(\bar{k}_-, \bar{k}_+)\in SO(4)$ maps the vector $N^I\!=\!(1,0,0,0)$ to $k^I$, then $k\!=\!\bar{k}_+ \bar{k}_-^{- 1}\in SU(2)$. 
One can also change variables to $\bar{B} \!=\! B+ \frac{1}{\gamma}\!\ast\! B$ so that $\beta \bar{B}_+^i + k \bar{B}_-^i k^{-1} = 0$ with $\beta=\frac{\gamma -1}{\gamma +1}$. 
We define a TGFT field representing a quantum tetrahedron $\varphi_k(g_1, \cdots g_4)\leftrightarrow \varphi_k(x_1, \cdots x_4)$, with $g_i\in SO(4)$ and $x_i\in\mathfrak{so}(4)$ (representing triangle bivectors). On this, we impose: 
\begin{equation} \label{gaugeproj}
(C \triangleright \varphi)_k = \int d h \, E_h \cdots E_h \star \varphi_{h^{-1} \triangleright k} \qquad .
\end{equation}
Upon integration over $k$, this gives the closure condition (this is a generalization of the invariance of the Boulatov TGFT field). Then, we impose the simplicity condition using the function: 
\begin{equation} \label{simpfunct}
S_k^\beta(x) := \delta_{- kx^- k^{-1}}(\beta x^+) = \int_{SU(2)} \!\! d u \, e^{i  tr[k^{-1} u k x^-]} e^{i \beta tr[u x^+]}
\end{equation} 
where $\delta_{- a}(b)\!:=\!\delta(a+b)$ is the $\mathfrak{su}(2)$ non-commutative delta function, as:
\begin{equation} \label{simplicity}
(S^\beta \triangleright \varphi)_k(x_1, \cdots, x_4) = \prod_{j=1}^4 S_k^\beta(x_j) \star \varphi_k (x_1,\dots x_4) \qquad .
\end{equation}
Defining $\widehat{\Psi}_k\!:=\!  S\triangleright C\triangleright\varphi_k= C\triangleright S\triangleright\varphi_k$, the action for the TGFT model imposing all the geometric conditions is 
 \begin{eqnarray} \label{actionbeta}
S \,\!\!=\!\!\,\frac{1}{2} \int [d^6 x_i]^4 \,d k  \, \varphi_{k1234} \star \varphi_{k 1  2 3 4} \nonumber  
+ \frac{\lambda}{5!}  \int [d^6 x_i][d k_a]   \,
\widehat{\Psi}_{1234k_a} \star \widehat{\Psi}_{4567k_b}\star  \widehat{\Psi}_{7389k_c}\star  \widehat{\Psi}_{962\,10k_d} \star \widehat{\Psi}_{10\,851k_e} 
\end{eqnarray}
where  the star product pairs repeated indices. 

The Feynman amplitudes of this TGFT take the form of a non-commutative simplicial path integral for the Holst-Plebanski action (with Immirzi parameter $\gamma$), and a quantum measure including geometric constraints on both bivectors and discrete connection. The expansion of the same amplitudes in group representations gives a spin foam model, encoding these constraints as conditions on the embedding of $SU(2)$ representations into $SO(4)$ representations. See also \cite{GFT-NC-review} for more quantum gravity applications of the non-commutative flux representation.

\section{Why adding pre-geometric information}
Let us now summarize some motivations for enriching tensor models with pre-geometric data to give TGFTs. The first reason is exemplified by the TGFT presented above: due to the additional data, the Feynman amplitudes of TGFTs can be given by simplicial gravity path integrals and, dually, by spin foam models. Thus, TGFTs are a 2nd quantized formulation of the spin network dynamics of LQG and connect directly with other quantum gravity approaches. From them, they can import techniques and physical insights, which in turn can guide both TGFT model building and the analysis of the resulting models. Compared to tensor models, TGFTs can be endowed with new symmetries, e.g. the analogue of 3d simplicial diffeomorphisms \cite{GFTdiffeos}, possibly leading to different critical behaviour and phase structure. A true renormalization flow for TGFTs can be defined and the issue of renormalizability addressed \cite{GFTrenorm}. Last, the additional pre-geometric information is an important guide for \lq reading out\rq possible geometric and physical meaning from the TGFT field and action and states, using the insights coming from LQG, spin foam models and simplicial gravity. The possibility of such physically motivated guesses and approximations will be crucial in the analysis of any TGFT model for gravity, and in any attempt to relate it to effective continuum physics (e.g. via mean field methods \cite{GFTmean}).

\end{document}